\documentstyle[preprint,tighten,aps,floats,psfig,epsf]{revtex}

\newcommand{\be}{\begin{equation}}
\newcommand{\ee}{\end{equation}}
\newcommand{\ba}{\begin{eqnarray}}
\newcommand{\ea}{\end{eqnarray}}

\begin{document}

\preprint{TTP99-09, hep-ph/9902399}

\title{Measuring $\sigma(e^+ e^- \to {\rm hadrons})$ using tagged
photon}

\author{S. Binner, J.H. K\"uhn
and K. Melnikov \thanks{
e-mail:  melnikov@particle.physik.uni-karlsruhe.de}}
\address{Institut f\"{u}r Theoretische Teilchenphysik,\\
Universit\"{a}t Karlsruhe,
D--76128 Karlsruhe, Germany}
\maketitle

\begin{abstract}
We propose to use events with radiated photons
in $e^+e^-$ collisions to measure the total
cross section of $e^+ e^- \to {\rm hadrons}$
as a function of the center of mass energy.
The Monte Carlo simulation for the collider 
DAPHNE shows that a competitive accuracy can be
achieved with this method. 

\end{abstract}
                       
\section{Introduction}

During the past years electroweak precision measurements have become one
of the central issues in particle physics. The indirect determination of
the mass of the top quark through its impact on quantum corrections
prior to its observation in proton-antiproton collisions at the TEVATRON
can be considered as one of the triumphs of the present theoretical
framework. Similarly the indirect determination of the mass of the $W$
boson is in perfect agreement with the present measurements at LEP and
the TEVATRON, which motivates a combined fit to  the parameters of the
Standard Model. 

Recently the measurements have become sufficiently precise to even give
first indications for a relatively small mass of the Higgs boson with an
upper limit of around 200 GeV. Significantly improved measurements, in
particularly of $M_W$ and the left-right asymmetry with polarized beams
at the SLC are expected to come in  the next couple of years which 
might lead to a fairly precise indirect determination of the Higgs 
mass $M_H$, thus repeating the success of the top quark mass determination.

An important ingredient in all these fits is
the fine structure constant at the mass of the $Z$ boson.
Its running from the Thompson limit to the high scale $M_Z$ is largely
determined by the hadronic contribution to the vacuum
polarization which in turn can be expressed via dispersion relation by 
$\sigma _{\rm had}$, the cross section for
$e^+e^- \to ~{\rm hadrons}~$. At present the dominant uncertainty in
the SM fits is due to the fairly large experimental error in
$\sigma _{\rm had}$ if the analysis is based on data only. 
Also the theoretical interpretation of future,
improved determinations of the muon anomalous magnetic moment is affected
by the limited knowledge of the hadronic contributions.
Significant progress
can only be achieved through an improved determination of the cross
section, either through substitution of inadequate data by theoretical
predictions based on perturbative QCD, wherever applicable, or through
improved measurements of $\sigma _{\rm had}$ over a wide energy range, or by a
combination of both.

All present, data based evaluations make use of a large variety of
experiments at different accelerators, a consequence of the large energy
range to be spanned by the dispersion integral. Some of the data points,
e.g. around the $\rho$ resonance, are now extremely accurate, others,
e.g. in the region between 2 and 3 GeV have large errors and are only
marginally consistent with the predictions based on pQCD. While a highly
precise scan through the whole region would certainly be desirable, one
may, as a viable alternative, exploit the high luminosity of oncoming
$e^+e^-$ machines which operate at fixed energy and use the radiative
return to lower energies. This would lead to an improved determination
of $\sigma _{\rm had}$ over a large energy range. The feasibility of this
approach is the subject of this work.

Similar considerations can be found in \cite{ChenZerwas,Ditt,Kur,Spag}.
However, in most of these papers only initial state radiation is considered
which is a priori legitimate  for  very small angles. 
In fact, Ref. \cite{Kur} uses explicitly the small angle approximation and 
includes higher order corrections  within the same approximation. This is an
attractive option {\em if} photon detection at very small angles
is experimentally feasible. Unfortunately, this seems to be not the 
case for the larger part of the oncoming high luminosity 
$e^+ e^-$ machines. In the present paper we consider
photons at somewhat larger angles, $\theta_\gamma > 7^{\circ}$ and
include initial and final state radiation 
and incorporate in addition collinear
initial state radiation through the structure
function technique. 

For the present purpose it is important to separate contributions
from initial and final state radiation. Three techniques can 
be envisaged:

i) Initial state radiation strongly dominates in events with 
photons emitted at small  angles relative to the beam axis. Imposing at
the same time an angular separation between photon and pions
decreases the contribution of the final state radiation 
even further;

ii) the forward-backward asymmetry of the pions which originates from
the interference between initial and final state radiation can be
measured and used to test the underlying model for final state
radiation;

iii) the angular distribution of photons from final state
radiation involves terms proportional to 
$\sin^2 \theta_\gamma$ and $\cos^2 \theta_\gamma$
only and is distinctly different from the one of initial 
state radiation with its strong peaking at small angles. By defining
suitable projectors it may be possible to distinguish the two 
components.

As a particular example we consider here the case of collider 
DAPHNE at Frascati which would operate on $\phi$ resonance;
the total energy of the collision therefore being equal
$\sqrt{s}=1.02$ GeV. The physical program of KLOE collaboration
is focused on the precision study of the CP violation in kaon system.
However, the events with radiative return to lower energies 
will be in quantity there. These events can be either 
considered as a background to an interesting physics
or utilized to study the cross section  $e^+e^- \to {\rm hadrons}$ 
at lower energies. In what follows we show that the second
option appears to be rather realistic. The studies presented
here are based on the Monte Carlo event generator \cite{SA}
that has been written for this purpose
and which, in addition, should be useful for the analysis of the data. 

\section{Basics of theoretical consideration}

Before we start with a more detailed discussion of the 
two pion case at DAPHNE, let us present the order of
magnitude of the event rates for a few characteristic 
configurations. The differential cross section for 
radiative events with $\theta_{\rm min} < \theta < 180^\circ -
\theta_{\rm min}$ is given by\footnote{Only initial state radiation is
taken into account.}:
\be
Q^2 \frac {{\rm d}\sigma}{{\rm d} Q^2} = \frac
{4 \alpha^3}{3 s} R(Q^2)
\left \{ \frac {(s^2+Q^4)}{s(s-Q^2)} 
\log \frac {1+\cos \theta_{\rm min}}{1-\cos \theta_{\rm min}}
-\frac {(s-Q^2)}{s}\cos \theta_{\rm min} \right \}.
\label{1}
\ee
Here $Q^2$ is the invariant mass of the hadronic system and $R(Q^2)$ is 
$\sigma(e^+ e^- \to {\rm hadrons})/\sigma_{\rm point}$.
The value of $Q^2 = s-2\sqrt{s} \omega_\gamma$ is fixed by 
$\omega_\gamma$,  the energy of the photon.  It is important to
note that the function in the curly brackets in  Eq.(\ref{1}) is flat
for $Q^2 \ll s$.  This is similar to the weight function in the
dispersive  integral for $\alpha(M_Z)$ and should facilitate the 
experimental determination of this important quantity. We will return
to  this issue below.

As an illustrative example we consider the $\pi^+ \pi^- \gamma$
final state, a photon angular cut of $5^\circ,7^\circ,10^\circ$,
a realistic model for the pion form factor \cite{KS} and the 
typical parameters of four colliders: DAPHNE, $B$-factory, LEP
and LEP2.  The respective beam energies, the luminosities and the event 
rates as obtained using Eq.(\ref{1}), are listed in Table 1.

\begin{table}
\begin{center}
$$
\begin{array}{||c||c||c||c|c|c||}
\hline
&
&
&
\multicolumn{3}{|c|}{{\rm Event~rates}}\\
\hline
{\rm Collider }& \sqrt{s}
&{\rm Annual~luminosity}, {\rm fb}^{-1} & \theta_{\rm min} = 5^\circ 
& \theta_{\rm min} = 7^\circ   &  \theta_{\rm min} = 10^\circ 
\\ \hline \hline
{\rm DAPHNE} & 1.02 & 1.35 & 18\cdot 10^6 & 16 \cdot 10^6 
&  14 \cdot 10^6 \\ \hline \hline
B-{\rm factory} & 10.6 & 100 & 4 \cdot 10^6 & 3.5 \cdot 10^6 
&  3 \cdot 10^6 \\ \hline \hline
{\rm LEP1} & 92 & 0.24 & 125 & 109 
&  93 \\ \hline \hline
{\rm LEP2} & 183 & 0.2 & 27 & 24 
&  20 \\ 
\hline
\end{array}
$$
\vspace*{0.5cm}
\caption{Estimated number of  radiative 
events $e^+e^- \to \pi^+ \pi^- \gamma$ for different center of mass
energies from Eq.(\ref{1}). 
The cut on the photon energy is $0.1$ GeV. To calculate
the annual luminosity we used $1~{\rm year} = 10^7~{\rm sec}$. 
For LEP2 we used accumulated luminosity for the 1997 run with 
at $\sqrt{s}=183$ GeV.}
\end{center}
\end{table}

Another interesting class of events are those with the invariant mass
of the hadronic system between $1.5$ and $3.5$ GeV. The $R$-ratio in
this region is very poorly known and even a measurement with 
an accuracy of $5\%$ would be considered a significant improvement.
On one hand, no collider will operate in this region in the 
near future and, on the other hand, the $R$ ratio could be well determined
with this method at a $B$-factory. In Table 2 we list typical event 
rates, adopting for simplicity a constant $R$ value of $2$.

\begin{table}
\begin{center}
$$
\begin{array}{||c||c||c||c||}
\hline
&
\theta_{\rm min} = 5^\circ 
&
\theta_{\rm min} = 7^\circ 
&
\theta_{\rm min} = 10^\circ 
\\ \hline \hline
1.5 \le \sqrt{Q^2} \le 2& 11\cdot 10^5 &9.9 \cdot 10^5 & 8.4 \cdot 10^5 \\ \hline \hline
2 \le \sqrt{Q^2} \le 2.5&9 \cdot 10^5 &7.9 \cdot 10^5 & 6.7 \cdot 10^5 \\ \hline \hline
2.5 \le \sqrt{Q^2} \le 3&7.6 \cdot 10^5 &6.6 \cdot 10^5 & 5.6 \cdot
10^5 \\ \hline \hline
3 \le \sqrt{Q^2} \le 3.5&6.7 \cdot 10^5 &5.8 \cdot 10^5 & 5.0 \cdot 10^5 \\ 
\hline
\end{array}
$$
\vspace*{0.5cm}
\caption{Estimated number of  radiative 
events $e^+e^- \to {\rm hadrons}+\gamma$ at $B$-factories for 
one year of running ($\sqrt{s}=10.6~{\rm GeV}$, integrated
luminosity $100~{\rm fb}^{-1}$). We use $R = 2$ in the 
interval $1.5 \le \sqrt{Q^2} \le 3.5$. }
\end{center}
\end{table}

It is clear from these tables that large event rates are at hand, even
for severe cuts, which confine the photon to the region of relatively 
large angles. The main issues are the control of  final 
state radiation and  radiative corrections. These questions 
will now be studied in more detail for the $\pi^+ \pi^- \gamma$ 
state and for the KLOE experiment.

For $\sqrt{s} \le 1$ GeV the dominant hadronic final
state produced in $e^+e^-$ annihilation consists 
of a pair of charged pions. For this reason 
we limit the discussion to  the reaction 
$e^+e^- \to \pi^+ \pi^- \gamma$ and demonstrate below
that by measuring the momenta of the
final state particles at fixed center of mass energy, 
one can extract accurate information about the cross 
section of $e^+ e^- \to \pi^+ \pi^-$ over a wide range 
of energy.

In the reaction  $e^+e^- \to \pi^+ \pi^- \gamma$ the photon 
can be emitted either by the electron or positron
(ISR) or by the pions in the final state (FSR). The corresponding
amplitude reads:
\be
{\cal M} = 
{\cal M}_{\rm ISR}+
{\cal M}_{\rm FSR}.
\ee

The calculation of  ISR is a straightforward application of the
Feynman rules for QED. In contrast, with the pions not
being elementary, the calculation of   FSR is in general
more tricky. In the current version of our Monte Carlo program,
we  consider pions as point-like particles 
as far as FSR is concerned  and use the standard 
Feynman rules of scalar QED. Being definitely not rigorous, 
this approach should provide a reasonable estimate of FSR
for photons emitted at small angles relative to  pions 
and an upper limit for photon emission at large angles.
Moreover, below we  present a useful tool
to control the accuracy of this approximation. Final state
radiation through the radiative decay $\phi \to \pi^+ \pi^- \gamma$
is not included in the analysis (for a recent discussion 
see \cite{Achasov} and references therein).
It can be considered as a special case of  FSR and 
we expect that this effect
can also be controlled through its interference with the dominant 
amplitude.

The interaction of the virtual photon with the $\pi^+ \pi^-$
system is described by a pion form factor $F_\pi(Q^2)$,
where $Q^2$ is the virtuality of the photon. A convenient
parameterization of this form factor was proposed in Ref.\cite{KS}.
The pion form factor is described by a sum of the Breit-Wigner
resonances with a $Q^2$ dependent width:
\begin{eqnarray}
&&F_\pi(Q^2) = \frac {BW_\rho \frac {(1+\alpha BW_\omega)}{1+\alpha}
+\beta BW_{\rho '}+\gamma BW_{\rho ''}}{1+\beta + \gamma};
\nonumber \\
&&BW_{\rho}(Q^2) = \frac {m_\rho^2}{m_\rho^2-Q^2-i\sqrt{Q^2}\Gamma_\rho(Q^2)}.
\end{eqnarray}
In the above equation 
$\alpha,~\beta,~\gamma$ are the  parameters of the model
(see Ref. \cite{KS}). For $Q^2=0$ one obtains $F_\pi(0)=1$; 
that ensures the  proper charge of a pion. In what follows 
we shall use this parameterization.

In general, by measuring the photon energy $\omega_\gamma$ 
one determines the invariant mass of the pions.
Consider now the square of the ISR amplitude.
In this case the differential cross section is evidently
proportional to the pion form factor squared at the proper
momentum scale:
\be
\left (\frac {{\rm d}{\sigma}}{{\rm d} Q^2} \right )_{\rm ISR}
\sim |F_\pi(Q^2)|^2 \sim \sigma_{e^+e^- \to \pi^+ \pi^-}(Q^2).
\ee

A different  situation occurs  when the 
photon is emitted from the final state. In this case
the differential cross section is proportional
to
\be
\left ( \frac {{\rm d}{\sigma}}{{\rm d} Q^2} \right )_{\rm FSR} 
\sim |F_\pi(s)|^2 \ne \sigma_{e^+e^- \to \pi^+ \pi^-}(Q^2).
\ee

Clearly, in this case measuring the energy of the final state photon 
would not help in determining the proper energy of the collision
and therefore the cross section of $e^+ e^- \to \pi^+ \pi^-$.
For this reason  FSR and ISR/FSR interference 
should be regarded as a ``background'' and the contribution 
due to  ISR as a ``signal''. It is definitely  not possible
to  ``switch off''  FSR completely , hence  FSR and the interference 
will lead to a systematic error in the determination of 
$\sigma (e^+ e^- \to \pi^+ \pi^-)$ from 
$\sigma (e^+ e^- \to \pi^+ \pi^- \gamma)$.
The choice of a proper selection criterion is thus 
the only possibility to suppress the contribution of 
FSR. In fact, the DAPHNE luminosity is so high, that  
even after imposing quite severe cuts, 
the number of radiative events remains huge, 
rendering the statistical accuracy of the measurement
a minor problem.  For this reason, it is the level of suppression 
of  FSR and the control of the remainder
that determines the accuracy of the 
measurement.  For the $\pi^+\pi^-$ final state, recent 
measurements in Novosibirsk have pushed the uncertainty in the pion
form factor down to one per cent for $\sqrt{s} < 1$ GeV
\cite{Eidelman}. Hence, for the proposed method to be competitive, 
the level of the FSR  contribution should not exceed this number.

In addition the FSR amplitude can be calibrated experimentally by
measuring the ISR-FSR interference which is odd under charge
conjugation and thus gives rise to a relatively large forward-backward
asymmetry.

\begin{figure*}
  \begin{center}
    \leavevmode
    \epsfxsize=8.cm
    \epsffile[19 144 519 643]{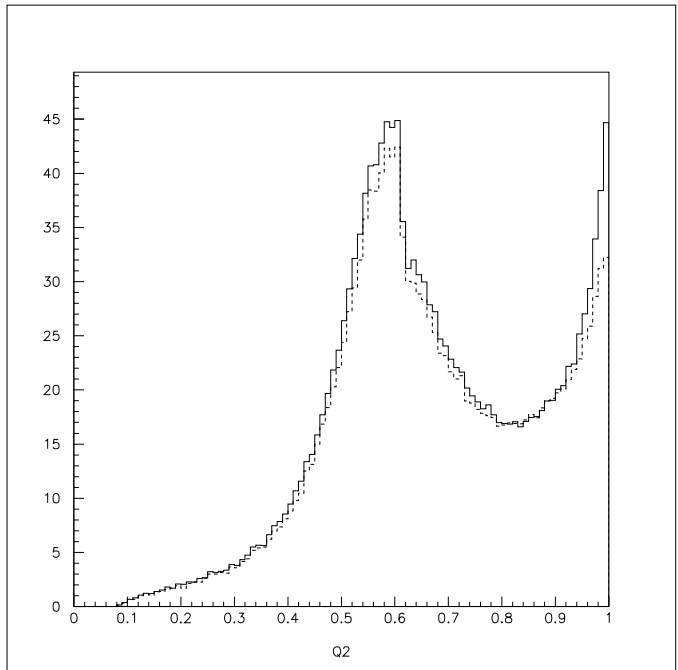}
    \hfill
    \parbox{14.cm}{
    \caption[]{\label{F1}\sloppy
The cross section 
${\rm d} \sigma(e^+e^- \to \pi^+ \pi^- \gamma )/{\rm d}Q_{\pi^+ \pi^-}^2$
in ${\rm nbarn}/{\rm GeV}^2$ as a function of $Q_{\pi^+ \pi^-}^2$ 
in ${\rm GeV}^2$ for
$\sqrt{s} = 1.02~{\rm GeV}$. The solid line is the distribution 
without collinear ISR, the dashed line -- with the collinear 
ISR. The cuts are $7^{\circ} < \theta_\gamma < 173^{\circ}$,
$ \omega_\gamma > 0.02~{\rm GeV}$ and the invariant mass 
of the detected particles in the final state 
$Q_{\pi^+ \pi^- \gamma}^2 > 0.9~{\rm GeV}^2$.}}
  \end{center}
\end{figure*}

In practice, the situation is further  complicated by additional
collinear initial state radiation which changes
the total energy of the collision and can not be detected.
This effect must be taken into account in any realistic Monte Carlo
event generator. As usual, the collinear initial state radiation  
is described by
the structure functions which provide a resummation
of the logarithmic corrections 
${\cal O}(\alpha \log ^2( s/m_e^2))$
to all orders in the coupling constant.  
We also assume that the invariant mass of the 
$\pi^+ \pi^- \gamma$ system can be measured 
sufficiently accurate. Events
with the invariant mass of the final state much
smaller than the total energy of the collision squared
can thus be rejected without loosing too much luminosity.
This allows to reduce significantly 
the kinematic distortion of the events due to  collinear initial 
state radiation. In the Monte Carlo program this is realized by
requiring a minimal invariant mass of the final state $\pi^+ \pi^- \gamma$.
For the structure functions the formulas of Ref. \cite{Rem}
are used.  For the purpose of  illustration, we present 
the results for the  pion invariant mass distribution with and 
without collinear initial state radiation in Fig.\ref{F1}.

In what follows we discuss two issues essential for a 
successful measurement. First, we  show that by choosing a proper
cuts one can significantly suppress the contribution due
to FSR. Second, we demonstrate that by measuring the 
forward-backward asymmetry of the produced pions one
can check how realistic the final 
state radiation amplitude is described.  Confronting our predictions
for the forward-backward asymmetry with the measurements,
one can gain  confidence in the predictions based on 
our Monte Carlo event generator.  If not stated otherwise,
the additional collinear radiation is incorporated with 
the structure function  technique.

\section {ISR dominance}

\begin{figure*}
  \begin{center}
    \leavevmode
    \epsfxsize=8.cm
    \epsffile[19 144 519 643]{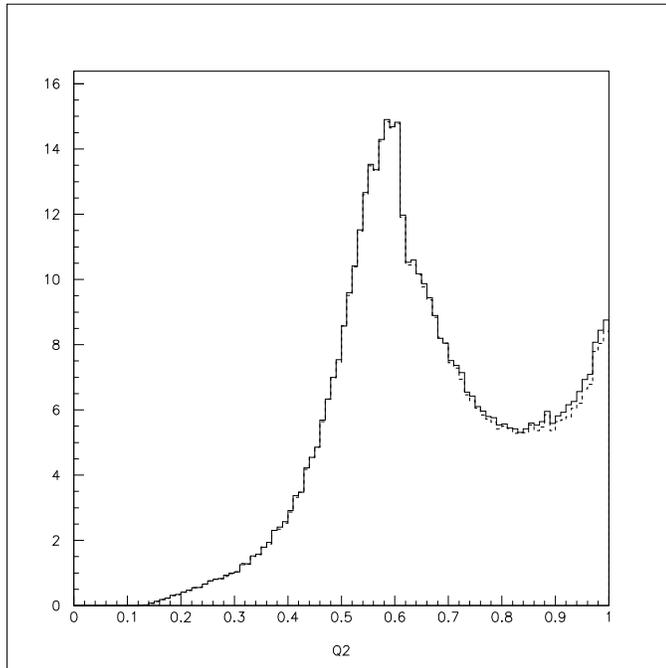}
    \hfill
    \parbox{14.cm}{
    \caption[]{\label{F2}\sloppy
A comparison of the ISR contribution (dashed line) with the
complete result (solid line). Plotted is  the cross section  
${\rm d} \sigma(e^+e^- \to \pi^+ \pi^- \gamma )/{\rm d}Q_{\pi^+ \pi^-}^2$
in ${\rm nbarn}/{\rm GeV}^2$ as a function of $Q_{\pi^+ \pi^-}^2$ 
in ${\rm GeV}^2$ for $\sqrt{s} = 1.02~{\rm GeV}$.  
The cuts are $7^{\circ} < \theta_\gamma < 20^{\circ}$
or $ 160^{\circ} < \theta_\gamma  < 183^{\circ}$,
$ 30^{\circ} < \theta_\pi < 150^{\circ}$,
$ \omega_\gamma > 0.02~{\rm GeV}$ and the invariant mass 
of the detected particles in the final state 
$Q_{\pi^+ \pi^- \gamma}^2 > 0.9~{\rm GeV}^2$.}}
  \end{center}
\end{figure*}

The spectrum of photons emitted by an ultrarelativistic particle
is described by the well-known formula:
$$
{\rm d} \sigma \propto {\rm d} \sigma_0 
\frac {{\rm d} \omega _\gamma}{\omega _\gamma} 
\frac { {\rm d}\theta _\gamma^2 }{\theta _\gamma^2}.
$$
Hence the relativistic particle strongly radiates in a small 
angular cone along its direction of motion. For this 
reason, a suppression of the FSR is achieved by selecting 
events where the pions are geometrically
separated from the hard photon. Simultaneously, an enhancement
of the ISR is achieved if one selects the events where the photon
is emitted at small angles relative to the collision axis. In reality,
the KLOE detector seems to be able to measure photons down
to $\theta_\gamma \sim 7^{\circ}$. Therefore, an appropriate
configuration is, for example, obtained by 
requiring $30^{\circ} < \theta_\pi < 150^{\circ}$
for pions and $7^{\circ} < \theta_\gamma < 20^{\circ}$ or
 $160^{\circ} < \theta_\gamma < 173^{\circ}$ for photons.
The prediction for the $Q_{\pi^+ \pi^-}^2$ invariant mass distribution 
is shown in Fig.\ref{F2}.  Note that we have included events with 
the photons emitted both in the forward and the backward
direction. Nevertheless, a forward backward asymmetry
of the $\pi^+$ (or, with opposite sign, of the  $\pi^-$) will be 
observed, which should 
not be confused with the trivial kinematic asymmetry if we would
consider photons with $7^{\circ} < \theta_\gamma < 20^{\circ}$
only.

It follows from Fig.\ref{F2} that for the chosen cuts the 
contribution of the FSR radiation to the total
cross section is of the order of $1$ per cent only. Also, 
one does not loose too much in the event rate since the 
pion angular distribution is relatively uniform and the 
photon angular distribution is strongly peaked at small angles.

Note that the contribution of the FSR to such a configuration 
comes only from the emission of photons by pions 
at {\em large} angles.  In this case the pions in the intermediate state
are far off-shell. For this reason, one expects that  
the point-like interaction of pions to photons is not likely to 
be an adequate description.
However, we believe, that in this case the point-like interaction 
{\em overestimates} the strength of the interaction and 
therefore gives a larger contribution due to the FSR.
Still, it is desirable to check how reasonable 
the interaction of photons to pions is described by our ansatz. 
A possible check is considered in the next Section.

\begin{figure*}
  \begin{center}
    \leavevmode
    \epsfxsize=8.cm
    \epsffile[19 144 519 643]{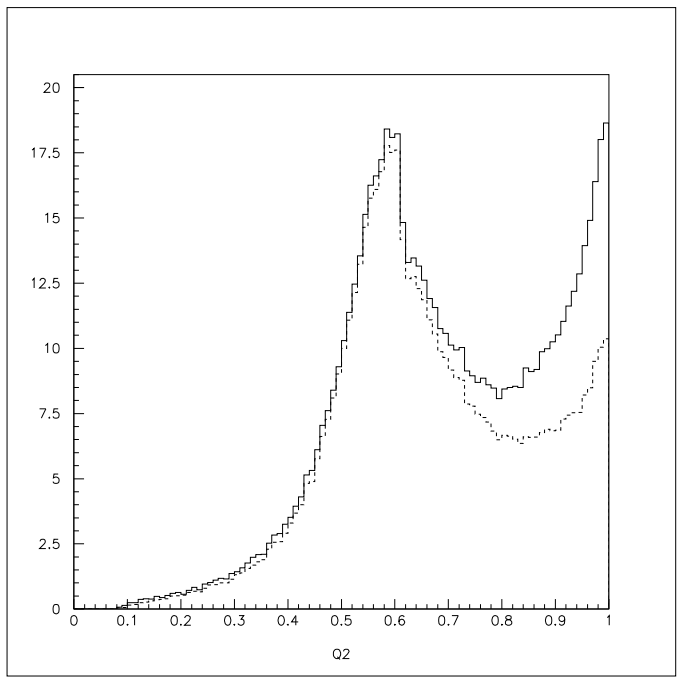}
    \hfill
    \parbox{14.cm}{
    \caption[]{\label{F3}\sloppy
The same as in Fig.\ref{F2}, but for angular cuts
$ 30^{\circ} < \theta_\gamma, \theta_\pi < 150^{\circ}$.}}
  \end{center}
\end{figure*}

The distribution shown in Fig.\ref{F2} may be compared with the distribution
when no geometric separation is imposed on photons and pions Fig.\ref{F3}.
In this case, one finds a significant FSR contribution. 
An interesting fact to note, is that the FSR contribution 
decreases for smaller $Q_{\pi^+\pi^-}^2$. This implies that possibly better
results can be achieved if a combined 
$Q_{\pi^+ \pi^-}^2$ - $\cos \theta_\gamma$ cut
is applied. Large $Q^2$, on the other hand, correspond to soft
photons where our description of  FSR is expected to work reasonably
well.

Let us imagine that the application of the proper selection 
criteria resulted in a suppression of the final state radiation.
In this case, even measuring the total cross section integrated over
the invariant masses of pions could provide  useful information 
for the determination of the fine structure constant. 

To illustrate this point, we consider the expression for the cross
section due to ISR:
\be
 \frac {{\rm d}\sigma_{ee \to \pi \pi \gamma}}
{{\rm d}Q^2 {\rm d} \cos \theta _\gamma} = \frac {\alpha^3}{3s^2}
|F_\pi(Q)|^2 \left ( 1- \frac {4m_\pi^2}{Q^2} \right )^{3/2}
 \left \{
\frac {(s^2+Q^4)}{Q^2(s-Q^2)}\frac {1}{1-\cos^2 \theta_\gamma}
-\frac {(s-Q^2)}{2Q^2} \right \}.
\label{eq6}
\ee

Consider now the case $s \gg Q^2$, i.e. a creation  of the 
pion pair with the invariant mass much smaller than the
energy of the collision. The above equation is then
simplified:
\be
 \frac {{\rm d}\sigma_{ee \to \pi \pi \gamma}}
{{\rm d}Q^2 {\rm d} \cos \theta _\gamma} \to
\frac {\alpha}{2\pi s} \sigma _{ee \to \pi \pi}(Q^2) 
\frac {(1+\cos \theta_\gamma^2)}{(1-\cos \theta_\gamma^2)}.
\ee

The total cross section for $e^+e^- \to \pi^+ \pi^- \gamma$ is 
obtained if one integrates over the invariant mass of pions and the 
photon production angle:
\be
\sigma_{ee \to \pi \pi \gamma} = 
\int \limits_{Q_{\rm min}^2}^{Q_{\rm max}^2} {\rm d}Q^2
\int  \limits_{\cos(\theta)_{\rm min}}^{\cos {(\theta)}_{\rm max}} 
{\rm d}\cos \theta_\gamma \frac {{\rm d}\sigma_{ee \to \pi \pi \gamma}}
{{\rm d}Q^2 {\rm d} \cos \theta _\gamma} \Rightarrow
\frac {\alpha}{4\pi s} {\cal F}(\cos(\theta)_{\rm max,min})
\int \limits_{Q_{\rm min}^2}^{Q_{\rm max}^2} {\rm d}Q^2 \sigma_{ee \to
\pi \pi} (Q^2). 
\ee

This expression should be compared with the small $s$ contribution 
to $\alpha_{\rm had}$.  Considering the contribution of the low-energy
region, $s \ll M_Z^2$, one obtains: 
$$
\delta \alpha_{\rm had}(M_z)  \approx \frac {\alpha(0)}{3\pi} 
\int \limits_{4m_\pi^2}^{s_{\rm max}} 
{\rm d}s\; \sigma_{ee \to {\rm hadrons}}(s) \propto 
\sigma_{e^+e^- \to \pi^+ \pi^- \gamma}.
$$

We therefore conclude, that even the integrated cross section 
for $e^+e^- \to {\rm hadrons} + \gamma$ can be effectively used
to obtain the contribution of the low energy region to the value
of the fine structure constant at the $Z$ resonance, provided
that the suppression of the final state radiation is achieved 
using appropriate selection criteria.

In a realistic situation, the accurate determination  
of  $\sigma(e^+ e^- \to \pi^+ \pi^- )$ from 
$\sigma(e^+ e^- \to \pi^+ \pi^- \gamma )$ requires 
the knowledge of the contribution due to  FSR and the effect of the 
collinear ISR. As we have explained above, FSR can be significantly
suppressed using  appropriate cuts on the photon and 
pion emission angles. After this is achieved, an appropriate
strategy which can be used to extract the pion form factor 
would be to fit the parameters of the model for the pion form factor
described in \cite{KS} to an observed $Q_{\pi^+ \pi^-}^2$-distribution 
using our Monte Carlo event generator which incorporates all the 
effects related to the collinear initial state radiation.

\begin{figure*}
  \begin{center}
    \leavevmode
    \epsfxsize=8.cm
    \epsffile[19 144 519 643]{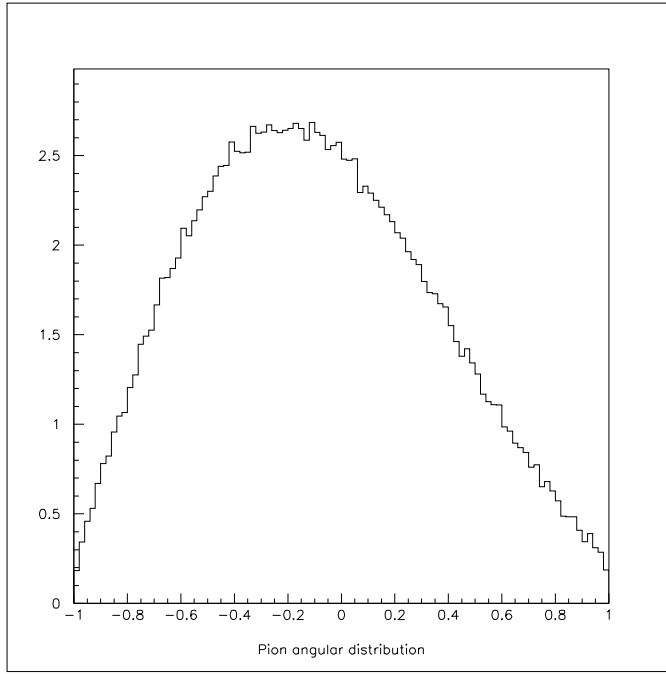}
    \hfill
    \parbox{14.cm}{
    \caption[]{\label{F4}\sloppy
Pion ($\pi^+$) angular distribution in the laboratory frame.
The cut on the photon angle is $ 60^{\circ} < \theta_\gamma  <
    120^{\circ}$.}}
  \end{center}
\end{figure*}

\begin{figure*}
  \begin{center}
    \leavevmode
    \epsfxsize=8.cm
    \epsffile[19 144 519 643]{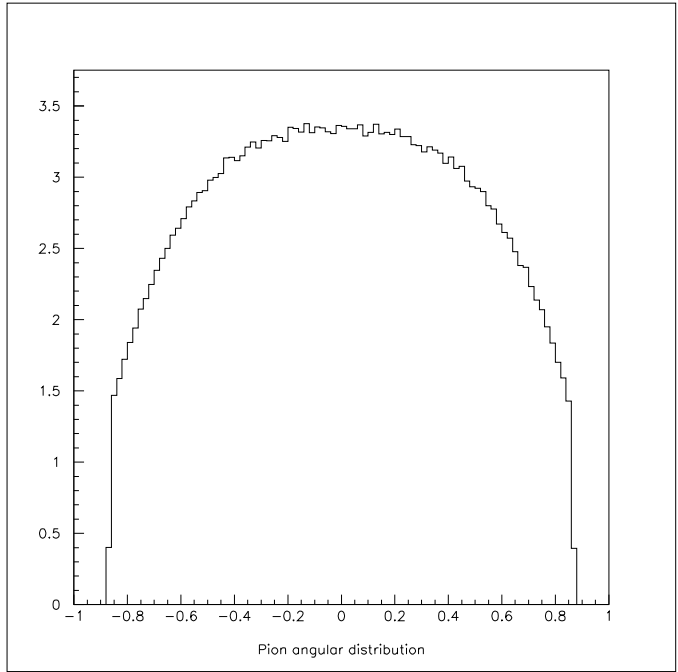}
    \hfill
    \parbox{14.cm}{
    \caption[]{\label{F5}\sloppy
Pion ($\pi^+$) angular distribution in the laboratory frame.
The photon angle is restricted to
$ 7^{\circ} < \theta_\gamma  < 20^{\circ}$ 
or  $ 160^{\circ} < \theta_\gamma  < 183^{\circ}$.
The cut on the pion angle is 
$30^{\circ} < \theta_{\pi^+,\pi^-} < 150^\circ$.
}}
  \end{center}
\end{figure*}

\section {Testing the model for the FSR}

The cross section for $e^+ e^- \to \pi^+\pi^-\gamma$
can be written as:
\be
{\rm d} \sigma \propto |{\cal M}|^2 = 
|{\cal M}_{\rm ISR}|^2+
|{\cal M}_{\rm FSR}|^2+
2{\rm Re} \left [ {\cal M}_{\rm ISR}{\cal M}^*_{\rm FSR} \right ].
\label{int}
\ee
We have explicitly separated the contribution due
to ISR, FSR and the interference of the two amplitudes.

If the photon is emitted from the initial(final) state, 
the pion pair is produced  with  charge parity $C=-1(+1)$.
Therefore,  the contribution  of the third term in Eq.(\ref{int}) 
vanishes if one integrates over the kinematic variables of the pions.
However, this term causes a significant 
charge asymmetry and, correspondently, a  forward-backward asymmetry
of $\pi^+$ and $\pi^-$.

As usual,  the asymmetry is defined  as:
\be
A = \frac {\int \limits_{-1}^{0} \frac {{\rm d}\sigma}
{{\rm d}\cos \theta _{\pi^+}} {\rm d} \cos \theta_{\pi^+}
-\int \limits_{0}^{1} \frac {{\rm d}\sigma}{{\rm d}
\cos \theta_{\pi^+}}
 {\rm d} \cos \theta _{\pi^+}
}
{\int \limits_{-1}^{0} \frac {{\rm d}\sigma}{{\rm d}\cos \theta _{\pi^+}}
 {\rm d} \cos \theta_{\pi^+}
+\int \limits_{0}^{1} \frac {{\rm d}\sigma}{{\rm d}\cos \theta _{\pi^+}}
 {\rm d} \cos \theta_{\pi^+}}.
\label{asy}
\ee
The asymmetry is also a function of the cuts on the photon angle
and energy which is implicitly assumed in the above equation.
This asymmetry should not be confused with the trivial 
asymmetry from kinematics which arises if one selects
photons in one hemisphere only.

For symmetric cuts on the photon angular distribution,
the numerator of this expression is non zero only due to the
interference term in Eq.(\ref{int}), the denominator 
equals to the total cross section. 
Given the dominance of the
initial state radiation in the total cross section, the 
forward-backward asymmetry is (roughly) proportional to the ratio
${\cal M}_{\rm FSR}/{\cal M}_{\rm ISR}$ and thus is 
a direct measure of the final state radiation.  For this reason, 
measuring the asymmetry would give an effective control over 
the accuracy of the  ansatz used to describe the 
interaction of photons to pions.

To demonstrate the magnitude of the effect consider the situation, where
the standard cuts on the photon energy and the invariant mass 
of $\pi^+\pi^-\gamma$ are applied. The photon angular cuts 
are such that $60^\circ \le \theta_\gamma \le 120^\circ$,
also $\omega_\gamma > 0.02$ GeV.
The angular distribution for positively charged pions is then
presented in Fig.\ref{F4}. One sees, that the distribution is not
symmetric: positively charged pions are mainly emitted in the 
backward direction. Using Eq.(\ref{asy}), we then obtain that the 
asymmetry $A$ equals  $20\%$. The total cross section in this 
case equals $3.29$ nbarn  which implies that up to $3.29\cdot 10^6$
events will be observed in a year of running.  Without any doubt, 
the twenty percent effect will be clearly visible with such a statistic.

Alternatively, we present in Fig.\ref{F5} the angular distribution
of pions for the case where the photon is emitted at small angles,
well separated from pions: $7^{\circ} < \theta_\gamma < 20^{\circ}$
or $160^\circ < \theta_\gamma < 183^\circ$, 
$30^{\circ} < \theta_{\pi^+,\pi^-} < 150^{\circ}$ and 
$\omega_\gamma > 0.02$ GeV. In this case the asymmetry 
is reduced to $A=1.5 \%$, demonstrating again the smallness
of final state radiation for this case.  

The separation between
initial and final state radiation is also possible for the symmetric
piece on the basis of the distinctly different angular distributions.
Let us for the moment ignore the folding with the radiator functions
and consider strictly the reaction $e^+ e^- \to \pi^+ \pi^- \gamma$.
Final state radiation alone leads necessarily to a photon angular
distribution of the form $\alpha + \beta \cos^2\theta_\gamma$,
initial state radiation on the other hand is described by 
Eq.(\ref{eq6}).  Given sufficient statistics, it is clear that 
a combined fit will allow  the separation of the two components.

\section {Conclusions}

In this paper we have argued, that one can effectively use 
the radiative return to low energies in order to measure
the cross section of $e^+ e^- \to {\rm hadrons}$ at variable
center of mass energy, of course  lower than the energy of the colliding 
beams. We have studied the feasibility of this approach 
with  particular emphasis on the two pion final state
-- a typical situation for the DAPHNE ring.
We have produced a realistic tool for these studies, 
a Monte Carlo event generator, which provides an option for 
choosing the cuts and includes both initial, final and 
collinear initial radiation.

By using  special selection criteria and 
measuring the energies and momenta of the particles in the 
final state one can suppress the contribution of the final 
state radiation to the one per cent level.  We have also demonstrated
that there will be a sizable forward-backward asymmetry, 
which can be used to check the quality of our description of the final 
state radiation.

For the forward-backward asymmetry to be significant
and to allow for this check, one should keep the photon angle 
sufficiently large, otherwise the ratio ${\cal M}_{\rm ISR}/{\cal M}_{\rm FSR}$
becomes too small. The opposite requirement must be fulfilled
for the accurate extraction of $\sigma(e^+e^- \to {\rm hadrons})$,
as discussed in Sect.3.  For this reason, all events with 
photons appear to be useful for extracting interesting information.

The proposed technique was mainly discussed in connection with the
DAPHNE collider. We would like to stress, however, that the 
use of the radiation return to lower energies can be used in 
a more general context and the same technique should, in principle,
be applicable also for other ``particle factories''.

\section{Acknowledgments}
We are grateful to H. Czyz for a number of useful discussions,
for his help in checking the event generator and for 
supplying routines for collinear radiation.
We would like to thank  G. Cataldi, A. Denig and  W. Kluge  for 
discussions concerning the KLOE detector and various 
aspects of the experimental situation.
 
This work was supported in part by BMBF under grant number
BMBF-057KA92P, by Graduiertenkolleg 
``Elementarteilchenphysik an Beschleunigern'' at the 
University of Karlsruhe and the EURODAPHNE network
TMR project ERB4061PL970448.


\begin{thebibliography}{10}

\bibitem{ChenZerwas} M.S. Chen and P. Zerwas, Phys. Rev. {\bf D11}
(1975), 58.

\bibitem{Ditt} M.W. Krasny, W. Placzek, H. Spiesberger,
Zeit. f. Physik {\bf C53} (1992), 687.

\bibitem{Kur} A.~B. Arbuzov {\it et al.}, hep-ph/9804430.

\bibitem{Spag} S.~Spagnolo, preprint CERN-OPEN-98-012.

\bibitem{SA} S. Binner, Diploma thesis, TTP, Karlsurhe, 1998 (unpublished).

\bibitem{Achasov} N.N. Achasov, V.V. Gubin and E.P. Solodov,
Phys. Rev. {\bf D55} (1997), 2672.

\bibitem{KS} J.H. K\"uhn and A. Santamaria, Zeit. f. Physik {\bf C48}
(1990), 445.

\bibitem{Eidelman} S.I. Eidelman and V.N. Ivanchenko, talk 
given at the at International Conference TAU98, Santander, Spain,
September 1998; to be published in the Proceedings.

\bibitem{Rem} M. Caffo, H. Czyz and E. Remiddi, Nuovo Cim. {\bf 110A}
(1997), 515; Phys. Lett. {\bf B327} (1994), 369.

\end{thebibliography}
\end{document}